\documentclass[
 amsmath,amssymb,
 aps,
]{revtex4-2}

\usepackage{graphicx}
\usepackage{dcolumn}
\usepackage{bm}
\usepackage{float} 
\usepackage{hyperref}
\usepackage[mathlines]{lineno}

\begin{document}

\preprint{APS/123-QED}

\title{Experimental investigation of intermediate-dissipation range energy spectra in shear turbulence}%

\author{Dipendra Gupta}
 \email{dg535@cornell.edu}
 \author{Edmund T. Liu}
\author{Gregory P. Bewley}%
\affiliation{%
 Sibley School of Mechanical and Aerospace Engineering, Cornell University, Ithaca, NY 14850, USA\\}

\date{\today}

\begin{abstract}
The shape of the turbulent energy spectrum in the dissipation range, 
where viscous effects dominate, remains an open question despite decades of work. 
We report an experimental investigation of intermediate dissipation range energy spectra in turbulent shear layers at Taylor-scale Reynolds numbers, $Re_\lambda$, ranging from approximately 450 to 1500, which are among the highest achieved in shear flow experiments that resolved small scales. 
We generated turbulent shear layers in a wind tunnel and measured using nanoscale hot-wire probes with a sensing length $l_w  \approx (0.2-0.5)\eta$ that was smaller than the Kolmogorov scale $\eta$ at all $Re_\lambda$. 
The measurements resolved wavenumbers up to 
$k_{max} \eta$ $\approx 17$ at the lowest $Re_\lambda$ and $k_{max} \eta$ $\approx 1$ at the highest $Re_\lambda$, where $k_{max}$ is the highest resolved wave number. 
In the range $0.1 \lesssim k \eta \lesssim 0.5$, 
the spectra collapse onto a universal stretched-exponential form, $E(k\eta) \sim $ exp$(-k\eta)^{\gamma} $, with $\gamma \approx 0.5$
independent of $Re_\lambda$. 
This value of stretching exponent, $\gamma$, is consistent with recent empirical and computational studies. 
The Reynolds-number invariance of $\gamma$ is strong evidence for universal scaling in the intermediate dissipation range of high-Reynolds-number shear turbulence. 
\end{abstract}

\maketitle

\section{\label{sec:level1}Introduction}

Understanding the statistical structure of turbulence across length scales remains one of the central challenges of classical physics. 
A cornerstone of modern turbulence theory is the concept of scale separation at high Reynolds number, whereby energy injected at large scales cascades through an inertial range before being dissipated at the smallest scales by viscosity. 
While the large-scale dynamics depend on forcing and boundary conditions, and the smallest scales are governed by viscous friction, the extent to which small-scale statistics are universal at high Reynolds numbers continues to be an area of active investigation.

Kolmogorov’s 1941 similarity theory (K41) \cite{kolmogorov1941local} provides a quantitative framework for spectral universality, predicting that in the inertial subrange — where viscosity is negligible and dynamics are governed solely by the mean dissipation rate $\epsilon$ — the energy spectrum converges toward high Reynolds numbers onto the well-known form, $E(k)=C_k \epsilon^{2/3}k^{-5/3}$. 
This prediction is supported by experiments and Direct Numerical Simulations (DNS) across a broad range of flows (see, {\it e.g.}, Ref. \cite{Pope_2000}). 
In contrast, the spectral structure beyond the inertial range, particularly in the dissipation range where viscous effects dominate, remains comparatively less well understood, particularly at high Reynolds numbers in laboratory flows. 

In the dissipation range, corresponding to wavenumbers $k\eta \gtrsim 1$ where $\eta=(\nu^3/\epsilon)^{1/4}$ is Kolmogorov length scale, the spectrum decays rapidly. 
A widely proposed general form of the energy spectrum in the dissipation range is 

\begin{equation}
    E(k\eta) =C(k\eta)^\alpha \exp(-\beta (k\eta)^\gamma)
       \label{eqn:eqn1}
\end{equation}

Numerous theoretical, computational, and experimental studies have sought to determine the value and universality of the dissipation-range exponent, $\gamma$, with reported estimates spanning approximately $0.5 \lesssim \gamma \lesssim2$. 
Early analyses, based on Taylor-series expansions of the velocity field, predicted $\gamma = 2$ \cite{townsend1951fine, novikov1961energy}. 
This value was supported by Smith and Reynolds (1991), which reports that
$\gamma = 2$ provides the best fit to the experimental data \cite{smith1991dissipation}. 
However, Manley (1992) argues that the measurements of Smith and Reynolds (1991) did not extend sufficiently beyond $\eta$, introduces a correction for the cutoff wavenumber, and concludes that $\gamma =1$ provides a more consistent fit to the data than $\gamma =2$ \cite{manley1992dissipation}. 
Earlier phenomenological arguments also suggest an exponential decay of the energy spectrum in the dissipation range. 
For instance, Kraichnan (1959) uses the Direct Interaction Approximation to predict that $\gamma =1$ \cite{kraichnan1959structure}. 
This prediction was supported by subsequent theoretical analyses \cite{sirovich1994energy, foias1990empirical, sreenivasan1985fine}, 
by direct numerical simulations (DNS) \cite{schumacher2007sub, ishihara2005energy,martinez1997energy}, 
and laboratory experiments \cite{saddoughi1994local, she1993universal}. 
More recently, however, analytical solutions of the fixed-point nonperturbative renormalization group (NPRG) equations associated with the Navier–Stokes equations predict
$\gamma =2/3$ \cite{canet2017spatiotemporal}. 
This latter result is roughly supported by recent experimental measurements \cite{gorbunova2020analysis,debue2018experimental}. 

Although DNS is inherently limited in achievable Reynolds numbers and scale separation, 
DNS has substantially advanced our understanding of the fine-scale structure of the dissipation range \cite{buaria2020dissipation, gorbunova2020analysis, schumacher2007sub, ishihara2005energy, chen1993far}. 
A common feature of these investigations is the implicit assumption of a pure exponential decay with $\gamma=1$, and a primary attention devoted instead to determining the power law scaling exponent $\alpha$ and the prefactor $\beta$. 
Consequently, the universality — and even the precise value — of $\gamma$ itself has not been directly tested, and remains an open question. 

Khurshid {\it et al.} \cite{khurshid2018energy} contains a systematic examination of the value of $\gamma$, using well-resolved DNS data up to $Re_\lambda \approx 100$. 
The work demonstrates that a single pure exponential provides an adequate representation of the spectrum in the dissipation range only at relatively low Reynolds numbers ($Re_\lambda \approx 20$). 
At higher $Re_\lambda$, significant deviations emerge in the large-wavenumber region ($k \eta \gtrsim 4$), commonly referred to as the far-dissipation range (FDR). 
To account for this behavior, Khurshid {\it et al.} \cite{khurshid2018energy} proposes a phenomenological model based on a superposition of two exponential contributions: 
one with $\gamma < 1$ governing the near-dissipation regime, NDR ($0.1 \lesssim k\eta \lesssim 4$), 
and another with 
$\gamma > 1$ dominating the FDR. 
This decomposition suggests that the dissipation spectrum is not characterized by a single universal exponent, but instead reflects a scale-dependent transition within the dissipation range itself.
Complementing this perspective, Buaria and Sreenivasan (2020) analyzes DNS data spanning $140 \leq Re_\lambda \leq 1300$ and reports that the effective exponent 
$\gamma$ decreases monotonically with increasing Reynolds number — from $\gamma \approx 2/3$ at the lowest $Re_\lambda$ to $\gamma \approx 0.5$ at the highest \cite{buaria2020dissipation}. 
However, this behavior was identified over a restricted band of wavenumbers, $0.15 \lesssim k\eta \lesssim 0.5$, 
which Buaria and Sreenivasan \cite{buaria2020dissipation} terms the intermediate dissipation range (IDR). 
Taken together, these DNS studies suggest that the dissipation spectrum cannot be fully characterized by a single Reynolds-number-independent exponential form. 
Rather, both Reynolds number and wavenumber location within the dissipation range appear to play a decisive role in shaping the effective decay exponent. 
Moreover, these numerical studies focused on homogeneous isotropic turbulence, underscoring the need for a more unified framework for the dissipation regime in general flows including shear-dominated ones. 

Several experimental investigations have extracted the scaling exponent $\gamma$. 
Early measurements \cite{saddoughi1994local, she1993universal, sreenivasan1985fine} largely supported a simple exponential form consistent with 
$\gamma = 1$, in line with the classical theoretical expectations. 
By contrast, Smith and Reynolds~\cite{smith1991dissipation} reports that $\gamma = 2$ provides the most satisfactory representation of the data. 
This divergence of conclusions drawn from high-quality laboratory experiments underscores the sensitivity of the inferred exponent to both measurement fidelity and the spectral window over which fitting is performed. 

Recent advances in instrumentation and flow control have enabled renewed scrutiny of the near-dissipation regime. 
For instance, experiments in von K\'arm\'an swirling turbulence revealed a stretched-exponential decay with $\gamma \approx 2/3$, consistent with predictions from NPRG analyses \cite{debue2018experimental}. 
Similar behavior was reported in grid-generated turbulence \cite{gorbunova2020analysis}. 
In these two experimental studies, the functional form expressed as a stretched exponential appeared approximately invariant with respect to Reynolds number over the range investigated. 
Nevertheless, the specific value of $\gamma$ differed from that inferred in earlier laboratory studies.

A plausible source of these discrepancies lies in spatial resolution. 
In principle, quantitative statements about the dissipative spectrum require 
$l_w \ll \eta$, so that attenuation effects remain negligible up to $k \eta \gtrsim 1$, where $l_w$ is spatial resolution of the measurement. 
Yet in several canonical experiments the probe length was comparable to or larger than $\eta$. 
For example, Saddoughi and Veeravalli (1994) employed a minimum sensing length $l_w \approx 2\eta $ \cite{saddoughi1994local}. 
While such resolution is entirely adequate for establishing inertial-range scaling, it cannot rigorously resolve the spectral curvature intrinsic to the near-dissipation regime. 
Even modest spatial filtering preferentially suppresses high-wavenumber content, artificially steepening or flattening the inferred decay depending on the characteristics of the instrumentation and data processing.

Thus, experimental studies broadly agree that the dissipative spectrum exhibits either exponential or stretched-exponential decay, but the precise value — and possible universality — of the exponent $\gamma$ remains unsettled. 
The confluence of limited spatial resolution, finite probe frequency response, restricted spectral fitting windows, and Reynolds-number variability complicates direct comparison among experiments. 
A definitive characterization of the dissipation-range spectrum therefore demands measurements with demonstrably sub-Kolmogorov spatial resolution, if one is to disentangle genuine physical scaling from instrumental artifacts. 

Shear flows present an especially important test case. 
Unlike idealized homogeneous isotropic turbulence, turbulent shear layers exhibit persistent large-scale anisotropy and vortical structures that may influence small-scale statistics. 
Although small-scale universality is often assumed at sufficiently high Reynolds numbers, systematic experimental verification of dissipation-range scaling in shear flows has been lacking due to resolution constraints. 
Establishing whether the dissipation-range spectrum exhibits a Reynolds-number-independent functional form in such flows is therefore of fundamental importance to turbulence theory.

In the present work, we present a systematic experimental investigation of the intermediate-dissipation-range energy spectra in turbulent shear layers over a wide range of $Re_\lambda$ varying from approximately 450 to 1500. 
Using nanoscale hot-wire probes with a sensing length smaller than the Kolmogorov scale across all cases, we resolve the smallest dynamically relevant scales. 
We demonstrate that in the range $0.1 \lesssim k\eta \lesssim 0.5$, 
the spectra collapse onto a universal stretched-exponential form with exponent $\gamma \approx 0.5$, independent of Reynolds number. 
These findings provide direct experimental evidence for Reynolds-number-independent scaling in the intermediate dissipation range of high-Reynolds-number shear driven turbulence.

\section{Experiments}

To examine dissipation-range statistics in a canonical turbulent shear flow, we selected the planar mixing layer as the base configuration, owing to its well-defined mean shear and its sustained production of small-scale turbulence. 
Experiments were conducted in the Warhaft Wind and Turbulence Tunnel at Cornell University, which features a 0.91\,m $\times$ 0.91\,m cross-section, a 9.1\,m long test section, and a maximum free-stream velocity of approximately 20\,m/s \cite{yoon1990evolution}. 
Background turbulence was generated using a 0.083\,m $\times$ 0.083\,m passive grid mounted at the inlet. 
A planar mixing layer was produced by establishing parallel streams of unequal mean velocity: the upper half of the grid was covered with a stainless-steel wire cloth (opening size 0.14\,mm), thereby increasing the local blockage and generating a velocity differential of 6\,m/s between the upper (low-speed) and lower (high-speed) halves of the tunnel. 
The upper free-stream velocity was 10.5\,m/s. 
The two streams were separated by an acrylic splitter plate positioned downstream of the grid. 
Because the velocity asymmetry induces a lateral pressure imbalance that tends to deflect the shear layer toward the high-speed side prior to reaching the splitter plate, a flow straightener was installed between the grid and the splitter plate to suppress large-scale deflection and ensure planarity of the mixing interface. A detailed description of the mean flow is provided in Ref.~\cite{gupta2023experimental}. 
The present configuration yields a statistically stationary planar shear layer with sufficiently high small-scale activity to interrogate the near-dissipation regime under controlled laboratory conditions.

Velocity measurements were obtained using nanoscale single hot-wire probes operated with a DANTEC Dynamics StreamLine Pro constant-temperature anemometer (CTA). Nanoscale hot-wire probes have emerged as a powerful diagnostic for resolving the smallest dynamically relevant scales in high-Reynolds-number turbulence \cite{baradel2024micro,le2021fabrication,borisenkov2015multiarray,bodenschatz2014variable,hultmark2012turbulent,bailey2010turbulence}, particularly within the dissipation range \cite{gorbunova2020analysis}. By achieving sub-Kolmogorov sensing lengths, $l_w \lesssim O(0.1\eta)$, such probes mitigate the spatial filtering inherent to conventional hot-wires and enable direct access to spectral content beyond $k\eta\sim1$. Recent studies have demonstrated their capability to resolve the curvature and functional form of near-dissipation spectra \cite{gorbunova2020analysis}.

\begin{figure} [H]
 \centering
 \includegraphics[width=0.9\linewidth, trim={50 260 50 150},clip]{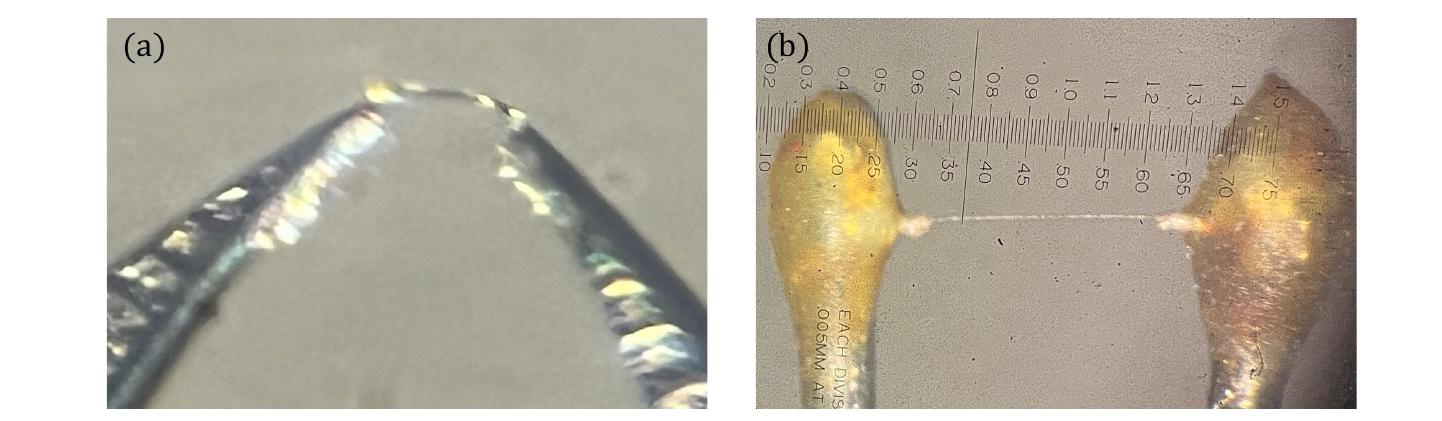}
    \caption{Hot-wire probes used in the present measurements. (a) Nanoscale hot-wire with active sensing length $l_w \approx 60,\mu\mathrm{m}$. \\ (b) Conventional hot-wire with $l_w \approx 1,\mathrm{mm}$.}
   \label{fig:Figure1}
\end{figure}
The nanoscale sensing element employed here was fabricated at the Cornell NanoScale Science and Technology Facility and had an active length $l_w \approx 60\,\mu\mathrm{m}$ and thickness $0.1\,\mu\mathrm{m}$ (see Fig.~\ref{fig:Figure1}a), providing sub-Kolmogorov spatial resolution over the range of conditions investigated \cite{liu2021nanoscale}. For comparison, we used conventional traditional platinum hot-wires ($l_w \approx 1\,\mathrm{mm}$ and diameter $\approx 5\,\mu\mathrm{m}$) (see Fig.~\ref{fig:Figure1}b) for one case in which it and the nanoscale probe both resolved the smallest scales, as a way to directly assess probe resolution effects. 
Both probes measured the longitudinal velocity component $u(t)$. 
Static calibrations were performed in the potential core of a round jet to ensure accurate voltage–velocity conversion. 
Analog signals were conditioned using a Krohn–Hite low-pass filter prior to digitization with a 16-bit National Instruments USB-6221 A/D converter at a sampling frequency of 40\,kHz, with record lengths exceeding $2.3 \times 10^7$ samples per realization. 

All measurements reported herein were acquired along the tunnel centerline at a downstream distance of 5\,m from the trailing edge of the splitter plate, where the mixing layer is well developed and in an approximately asymptotic state. 
Taylor’s frozen-turbulence hypothesis was invoked to convert temporal signals into spatial statistics, using the local mean velocity $\overline{u(t)}$ as the convection speed. 
The Kolmogorov length scale was computed as $\eta=(\nu^3/\epsilon)^{1/4}$ where $\epsilon$ is mean turbulent kinetic energy dissipation rate. 
To ensure robustness, $\epsilon$ was evaluated using three methods \cite{ schroder2024estimating,rusello2011turbulent}: 
(i) the isotropic surrogate based on filtered velocity gradients, $\epsilon = 15 \nu~\overline {({\partial u’}/{\partial x})^2}$, where $u'(t)= u(t) - \overline{u(t)}$ and $u(t)$ is instantaneous turbulent velocity; 
(ii) the second spectral moment of smoothed spectra, $\epsilon = 15 \nu\int k^2 E(k) dk$, where $k$ is wavenumber; and
(iii) the third-order longitudinal structure function in the inertial range via Kolmogorov’s four-fifths relation \cite{kolmogorov1941local,Pope_2000}. Among the three approaches, methods (i) and (ii) yield consistent estimates of the dissipation rate, agreeing within 3\% and systematically exceeding those obtained from method (iii) by as much as approximately 25-30\% at the highest Reynolds number investigated here. Recent study \cite{schroder2024estimating} has shown that the gradient-based estimate provides the most accurate determination of the dissipation rate compared to alternative approaches, including method (iii). Accordingly, we adopt the gradient-based estimate for computing $\eta$. This choice ensures that $\eta$ is not overestimated, thereby providing a conservative and reliable assessment of dissipative-range resolution.
The Taylor microscale was evaluated as $\lambda = [{2\overline{u’^2}}/{\overline{(\partial u'/\partial x)^2}}]^{1/2}$~\cite{Pope_2000} and the corresponding Reynolds number $Re_\lambda = \sqrt{\overline{u'^2}}\lambda/\nu$. 

Through the combined use of nanoscale hot-wire anemometry and conservative dissipation estimates, the present measurements were designed specifically to minimize instrumental contamination of the near-dissipation-range spectrum. 
This level of control is essential when interrogating subtle variations in the functional form and exponent of spectral decay at large wave numbers, where genuine physical curvature must be distinguished from artifacts of finite probe resolution and bandwidth limitations. 

\begin{figure} [H]
 \centering
 \includegraphics[width=0.9\linewidth, trim={50 58 50 130},clip]{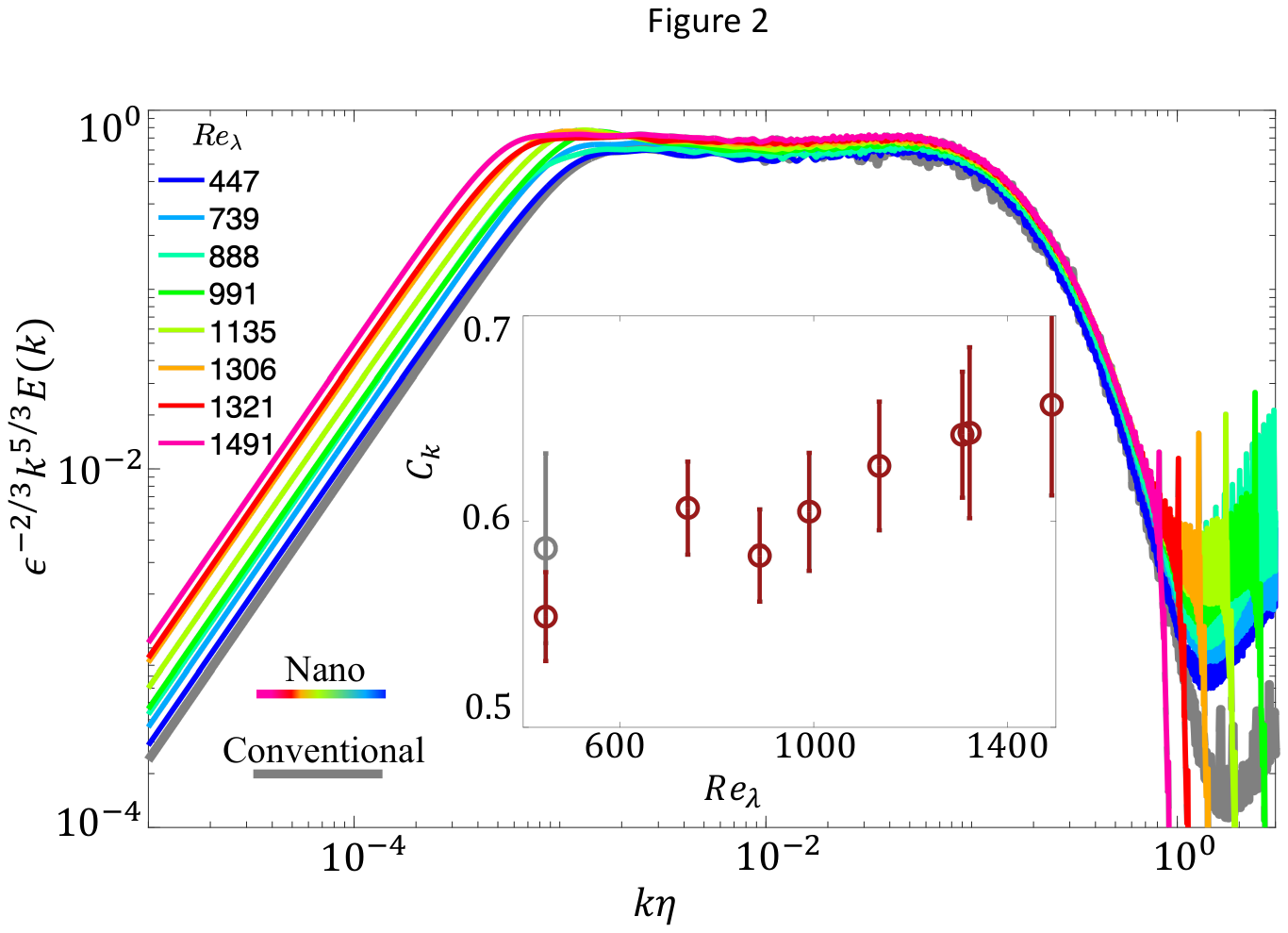}
    \caption{Compensated longitudinal energy spectra measured over a range of Taylor-scale Reynolds numbers, $Re_\lambda$. Spectra obtained using the nanoscale hot-wire probe are shown in color, while the spectrum measured with a conventional hot-wire at $Re_\lambda \approx 450$ is shown in gray. The agreement between the two probes extends up to $k\eta \approx 1$, indicating negligible spatial-resolution bias over the commonly resolved scales. {\it Inset:} Reynolds-number dependence of the Kolmogorov constant, $C_k$, estimated from the inertial-subrange plateau. Error bars represent the standard deviation computed over the identified plateau region for each dataset. Dark red circles denote estimates obtained from the nanoscale probe, while that from the conventional probe is in gray.}
   \label{fig:Figure2}
\end{figure}

\subsection{Compensated energy spectra measured with the nanoscale hot-wire at varying $Re_\lambda$ }

Figure \ref{fig:Figure2} presents the compensated longitudinal energy spectra, $E(k)k^{5/3}\epsilon^{-2/3}$, obtained using a nanoscale hot-wire over a broad range of $Re_\lambda$. To assess effects of spatial resolution, we first compare spectra at $Re_\lambda \approx 450$ obtained using both nanoscale and conventional hot-wire probes. The two datasets exhibit excellent agreement up to $k\eta \approx 1$, encompassing the inertial subrange and the onset of dissipation. In particular, both probes recover a well-defined inertial plateau with $C_k \approx 0.55$, consistent with established results \cite{sreenivasan1995universality}. This consistency demonstrates that the nanoscale probe does not introduce any measurable systematic bias over the range of scales that are commonly  resolved by both instruments..

When plotted as a function of the normalized wavenumber $k\eta$, the spectra obtained using the nanoscale probe exhibit the expected collapse in the inertial subrange across all investigated $Re_\lambda$.
Over nearly two decades in wavenumber, a well-defined plateau is observed with Kolmogorov constant $C_k = 0.62 \pm0.03$ (see Fig. \ref{fig:Figure2} {\it (inset)}). 
A closer examination reveals a modest but systematic increase in $C_k$ from approximately 0.55 at the lowest $Re_\lambda$ to about 0.65 at the highest values. 
Beyond $Re_\lambda \approx 1300$, the prefactor appears to saturate near 0.65, remaining within the range commonly reported in laboratory and numerical studies \cite{sreenivasan1995universality}. 
This gradual rise of $C_k$ with Reynolds number is consistent with the notion that inertial-range universality is approached asymptotically, with finite-$Re_\lambda$ corrections persisting even at the Reynolds numbers we reached.

A subtle bottleneck is discernible near $k\eta \approx0.02$, manifesting as a slight elevation of the compensated spectra above the inertial plateau prior to the onset of dissipative decay. 
This feature persists at all $Re_\lambda$ investigated, in agreement with previous experimental observations \cite{kuchler2019experimental}. 
The persistence of the bottleneck at high Reynolds number underscores its dynamical origin rather than being a finite-resolution artifact. 
The departure from inertial scaling—the spectral roll-off—commences at approximately $k\eta \approx0.08$, marking the beginning of the near-dissipation regime. Beyond this point, the compensated spectra progressively deviate from the plateau and enter a regime characterized by rapid decay. Notably, while the amplitude of the spectra exhibits a Reynolds-number dependence, their curvature and overall functional form remain remarkably similar. 
This suggests that the dissipative structure retains an approximately universal shape in normalized coordinates.

At the highest Reynolds numbers, the smaller Kolmogorov scales push the dissipative band toward increasingly higher frequencies. 
Consequently, measurement noise becomes non-negligible for scales larger than $k\eta \approx 1$. 
Although the nanoscale probe satisfies $l_w/\eta \ll 1$ for all cases considered, the combined effects of finite sampling frequency and electronic noise impose a practical upper limit on the reliably resolved wavenumber range. 
For this reason, quantitative analysis of the spectral decay in the dissipation range is conservatively restricted to $k\eta \approx0.5$, beyond which the highest-$Re_\lambda$ spectra exhibit incipient noise contamination. 
Restricting attention to this well-resolved band ensures that subsequent inference regarding the functional form of the near-dissipation decay—whether exponential or stretched exponential—is not influenced by instrumental limitations. 
The consistent inertial-range collapse, bottleneck behavior, and Reynolds-number-independent onset of roll-off collectively provide confidence that the nanoscale hot-wire captures the physically relevant curvature of the spectrum in the near-dissipation regime.


\section{Universality of the Spectral Decay in the Intermediate Dissipation Range}

Figure \ref{fig:Figure2} demonstrates that the spectra exhibit a remarkably similar decay beyond the inertial subrange, at least up to $k\eta \approx 0.5$, beyond which instrumental noise is non-negligible. 
The similar spectral shape in this near-dissipation interval strongly suggests the existence of an approximately universal functional form. 
The central question is therefore whether the exponent $\gamma$ in Eq.~\ref{eqn:eqn1} is itself universal, {\it i.e.}, independent of Reynolds number and large-scale flow conditions.
A convenient starting point is the logarithmic derivative of the spectrum, which has been widely employed in dissipative-range analyses \cite{buaria2020dissipation,khurshid2018energy,martinez1997energy}:
\begin{equation}
\phi(k\eta) \equiv \frac{d \log E(k\eta)}{d \log (k\eta)}
= \alpha - \beta \gamma (k\eta)^{\gamma}.
\label{eqn:eqn2}
\end{equation}
This formulation eliminates the multiplicative prefactor of $E(k\eta)$ and isolates the functional dependence of the decay. 
In principle, $\gamma$ can be obtained directly from this relation. 
In practice, however, the extraction is complicated by the nonlinear coupling of the three parameters $(\alpha,\beta,\gamma)$ and by the increasing sensitivity to noise as $k\eta$ grows.
Historically, many studies have assumed a purely exponential decay ($\gamma=1$) and focused on determining $\alpha$ and $\beta$ \cite{saddoughi1994local,martinez1997energy,ishihara2005energy,schumacher2007sub}. 
While this assumption simplifies the fitting procedure, it effectively suppresses any deviation from exponential scaling and can obscure values of $\gamma$. 
A more systematic treatment in Ref.~\cite{khurshid2018energy} reports $\alpha \approx 0$ in the near-dissipation range. 
This significantly reduces the dimensionality of the fitting problem, though it still leaves the coupled product $\beta\gamma$ unresolved. 
Ref.~\cite{khurshid2018energy} addresses this difficulty through a compensation procedure, seeking plateaus in $\phi(k\eta)[\gamma(k\eta)^{\gamma}]^{-1}$ for trial values of $\gamma$. 
Similarly, Buaria and Sreenivasan (2020) employs nonlinear least-squares fitting to determine $\beta$ and $\gamma$ independently \cite{buaria2020dissipation}. 
Alternative approaches based on higher-order logarithmic differentiation provide local estimates of $\gamma$ \cite{gorbunova2020analysis,debue2018experimental} but are notoriously sensitive to noise amplification, particularly at high Reynolds number where spectral resolution must extend deeper into the dissipation range.

In the present study, we adopt a strategy that balances robustness with minimal modeling assumptions. 
Guided by the findings in Refs.~\cite{buaria2020dissipation, khurshid2018energy}, we assume: (a) $\alpha = 0$ in the near-dissipation interval; (b) the product $\beta \gamma = c$, where $c$ is treated as an effective constant over the fitted range. Under these assumptions, the logarithmic derivative reduces to
$\phi(k\eta) = - c (k\eta)^{\gamma}$. The fitting problem thus involves only two parameters: $\gamma$ and $c$. 
Importantly, this form separates the curvature (controlled solely by $\gamma$) from the overall amplitude. 
Rather than performing nonlinear regression in linear space, we follow and refine the methodology in Ref. \cite{buaria2020dissipation} by recasting the fitting problem in logarithmic coordinates: 
$-\log \phi(x) = \log c + \gamma \log x$, with $x \equiv k\eta$. 
This transformation renders the estimation linear in parameters and is statistically consistent with the approximately multiplicative nature of spectral fluctuations. 
Moreover, it mitigates the scale-dependent weighting bias inherent in nonlinear least-squares procedures performed directly on $\phi(x)$. 
To quantify statistical uncertainty, the full spectral record was partitioned into $30$ statistically independent segments, each at least $50$–$100$ integral time scales in duration to ensure decorrelation. 
The parameters $(c,\gamma)$ were estimated independently for each segment. 
The standard error was obtained from ensemble-averaged values of the segment-wise estimates. 
This procedure reflects realization-to-realization variability and avoids underestimating uncertainty through overfitting of a single long record.

\begin{figure} [H]
 \centering
 \includegraphics[width=0.9\linewidth, trim={30 160 15 150},clip]{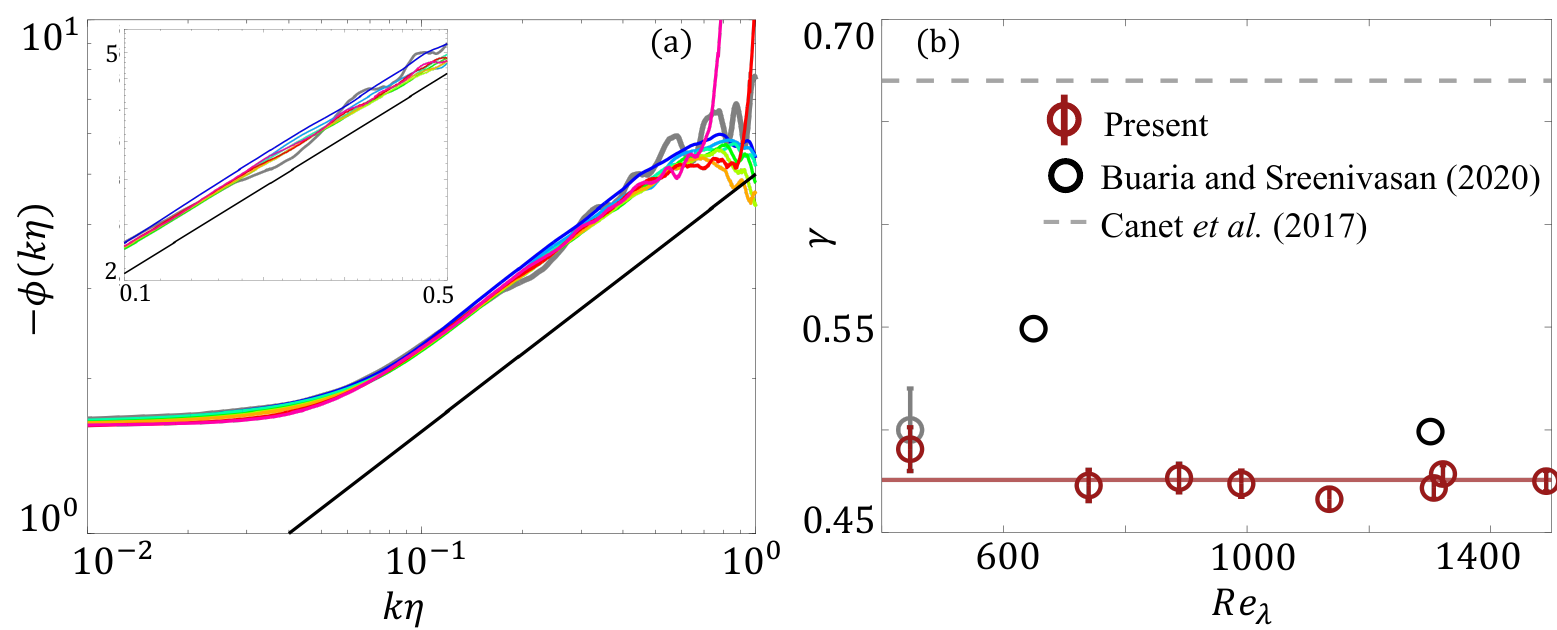}
    \caption{(a) Logarithmic derivative of the longitudinal energy spectrum, $d\log E(k)/d\log k$, plotted as a function of the normalized wavenumber $k\eta$ for different $Re_\lambda$. The black line denotes the reference scaling $(k\eta)^{0.5}$. {\it Inset:} Enlarged view over the interval $0.1 \leq k\eta \leq 0.5$. Colors are consistent with Fig.~\ref{fig:Figure2}. (b) Reynolds-number dependence of the intermediate-dissipation-range exponent, $\gamma$. Black circles correspond to data from Ref.~\cite{buaria2020dissipation}, and the gray dashed line indicates the theoretical prediction of \cite{canet2017spatiotemporal}. Dark red circles denote estimates obtained using the nanoscale probe, while gray circles correspond to measurements from the conventional probe. The solid dark red line represents the mean value of the nanoscale probe measurements.}
   \label{fig:Figure3}
\end{figure}

Figure \ref{fig:Figure3}a displays the logarithmic derivative $\phi(k\eta)$ as a function of $k\eta$ for the full range of $Re_\lambda$ investigated. 
Two dynamically distinct regimes are immediately evident. 
At low wavenumbers, the curves collapse to $\phi(k\eta)\approx -5/3$, consistent with Kolmogorov’s inertial-range prediction. 
This serves as an internal consistency check, confirming that the differentiation procedure preserves the inertial scaling and that the spectral estimates are sufficiently resolved in the inertial subrange. 
More importantly, in the interval $0.1 \lesssim k\eta \lesssim 0.5$, which we identify as IDR, the curves for all $Re_\lambda$ collapse remarkably well onto a common power-law trend. In logarithmic coordinates, this region exhibits a well-defined slope corresponding to a stretched-exponential exponent $\gamma \approx 0.5$. 
A quantitative least-squares regression, performed in logarithmic space as described above, yields $\gamma = 0.48\pm0.02$, 
as shown in Fig.~\ref{fig:Figure3}(b). 
Within statistical uncertainty, $\gamma$ is invariant with respect to $Re_\lambda$ over the entire range of Reynolds numbers considered here. 
No systematic drift or trend is observed. 
This invariance is a central result of our study.

Among existing studies at comparable Reynolds numbers, the DNS study in Ref.~\cite{buaria2020dissipation} provides the closest reference point. 
At the highest $Re_\lambda$, the reported $\gamma$ approaches values similar to ours. 
However, in Ref.~\cite{buaria2020dissipation} $\gamma$ exhibits a discernible Reynolds-number dependence over the accessible range. 
In contrast, within our statistical uncertainty, $\gamma$ remains constant across all $Re_\lambda$ examined.
Our findings are also consistent, within uncertainty, with the experimental study in Ref.~\cite{gorbunova2020analysis}, which reports $\gamma = 0.68\pm0.19$ in grid-generated turbulence with $60 \lesssim Re_\lambda \lesssim 500$. 
Although their central estimate is somewhat higher than ours, the relatively broad uncertainty interval encompasses ours. Importantly, they likewise observed no systematic Reynolds-number dependence of $\gamma$.
It is worth emphasizing that the fitting range used in Ref.~\cite{gorbunova2020analysis}, namely $0.2\lesssim k\eta \lesssim 3$,
extends well beyond the IDR interval identified here as exhibiting universal collapse. 
Our results indicate that the Reynolds-number–independent scaling is confined to a comparatively narrow band, approximately $0.1 \lesssim k\eta \lesssim 0.5$, similar to that observed in Ref.~\cite{buaria2020dissipation}. 
Beyond $k\eta \approx 0.5$, the present measurements cannot reliably resolve the decay due to increasing noise contamination, and we therefore refrain from making claims regarding universality deeper into the dissipation range.

The exponent we measured, $\gamma = 0.48\pm0.02$,
is lower than the value $\gamma = 2/3$ predicted by NPRG analyses developed under assumptions of statistical homogeneity and isotropy \cite{canet2017spatiotemporal}. 
Several factors may account for this. 
First, the present flow is a planar turbulent mixing layer, not homogeneous isotropic turbulence. Although small scales are expected to approach local isotropy at sufficiently high Reynolds numbers, residual anisotropy and nonlocal interactions associated with shear-layer dynamics may alter the curvature of the dissipative cutoff. 
Second, the exponent is sensitive to the fitting window; our analysis deliberately restricts attention to the interval $0.1 \lesssim k\eta \lesssim 0.5$, where collapse is demonstrably universal. 

More importantly, regardless of the precise numerical value, the observed Reynolds-number invariance of $\gamma$ is consistent with theoretical frameworks in which the near-dissipation range approaches an asymptotic regime characterized by a universal stretched-exponential form. 
In contrast, the multifractal model predicts a logarithmically varying effective exponent in the intermediate dissipation range \cite{frisch1991prediction}. 
Within the resolution of the present measurements, we find no evidence of such systematic variation, as also observed in Refs.~\cite{debue2018experimental, gorbunova2020analysis}. 
The data instead support convergence toward a constant exponent at sufficiently large Reynolds number. 
The discrepancy may reflect the fact that multifractal predictions correspond to asymptotic limits that may require Reynolds numbers beyond those presently accessible.

The intermediate-dissipation range represents a transitional regime where nonlinear transfer and viscous dissipation are comparable. 
The observed stretched-exponential decay with $\gamma \approx 1/2$ indicates a sub-exponential cutoff, reflecting gradually weakening nonlinear interactions rather than an abrupt viscous termination. 
In a planar mixing layer, residual shear and large-scale structures may sustain interactions into the dissipative scales, producing a softened roll-off. 
Remarkably, this curvature collapses across all $Re_\lambda$, demonstrating that $\gamma$ captures a universal feature of the viscous transition layer, largely insensitive to outer-scale anisotropy.

\begin{figure} [H]
 \centering
 \includegraphics[width=0.9\linewidth, trim={20 160 15 150},clip]{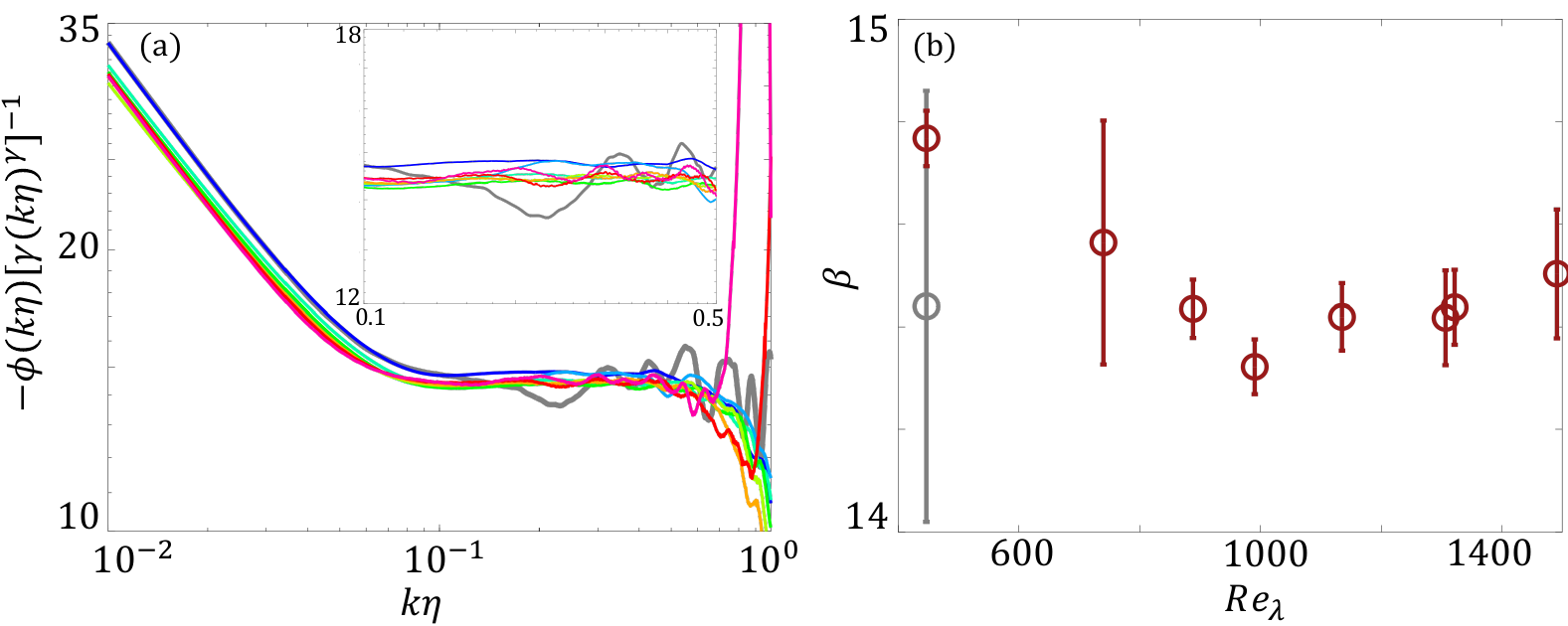}
    \caption{(a) Compensated logarithmic derivative of the longitudinal energy spectrum plotted as a function of the normalized wavenumber $k\eta$ for different $Re_\lambda$. {\it Inset:} Enlarged view over the interval $0.1 \leq k\eta \leq 0.5$. Colors are consistent with Fig.~\ref{fig:Figure2}. (b) Reynolds-number dependence of the exponent $\beta$. Dark red circles denote estimates obtained from the nanoscale hot-wire measurements, while gray circles correspond to measurements using the conventional probe.}
   \label{fig:Figure4}
\end{figure}

To further assess the robustness of the extracted exponent, we examine the compensated logarithmic derivative, $\phi(k\eta)/[\gamma (k\eta)^\gamma]$, which should yield a constant equal to $-\beta$ if $\alpha \approx 0$. 
As shown in Fig.~\ref{fig:Figure4}(a), all $Re_\lambda$ exhibit a clear plateau over $0.1 \lesssim k\eta \lesssim 0.5$, coinciding with the previously identified universal interval. 
The existence of this plateau independently confirms the negligibility of $\alpha$ in this regime and supports the internal consistency of the stretched-exponential representation. 
We attribute minor oscillations observed at the upper end of the interval, most pronounced at the highest $Re_\lambda$ and in measurements obtained with the conventional probe, to finite wavenumber resolution as  $k_{\max}\eta \to \eta/l_w$.
The plateau level directly yields $\beta$. 
Averaging over $0.1 \le k\eta \le 0.5$, we obtain $\beta = 14.5\pm0.3$, with no systematic dependence on $Re_\lambda$, as can be seen in Fig.~\ref{fig:Figure4}b. 
This value is consistent with the independent least-squares fits described earlier (not shown). Accordingly, $\beta\gamma \approx 6.9 \pm 0.2$, in close agreement with prior DNS and experimental studies \cite{khurshid2018energy, gorbunova2020analysis, debue2018experimental, saddoughi1994local}. 
The Reynolds-number invariance of both $\gamma$ and $\beta$ indicates that not only the curvature but also the amplitude of the near-dissipation spectral roll-off becomes universal, even in a strongly sheared planar mixing layer.

\section{Conclusion}

In this study, we have undertaken a systematic experimental investigation of the intermediate dissipation range in a planar turbulent mixing layer, employing nanoscale hot-wire probes capable of resolving sub-Kolmogorov scales at Taylor-scale Reynolds numbers $450 \lesssim Re_\lambda \lesssim 1500$. 
Within the interval $0.1 \lesssim k\eta \lesssim 0.5$, the energy spectrum is well described by a stretched-exponential form, $E(k\eta) \sim \exp(-\beta(k\eta)^\gamma)$,
with $\gamma = 0.48 \pm 0.02$ and $\beta =14.5\pm0.3$, both independent of $Re_\lambda$. 
Within the experimentally accessible window, the invariance of $\gamma$ is unambiguous and is in quantitative agreement with other DNS and experimental studies \cite{buaria2020dissipation, gorbunova2020analysis}. 
The absence of any systematic Reynolds-number dependence suggests that the intermediate-dissipation regime spectrum approaches a universal stretched-exponential curvature with $\gamma \approx 0.5$ in the IDR at sufficiently large Reynolds numbers. 
We emphasize, however, that the present measurements do not resolve the far-dissipation regime, where a transition toward a steeper or purely exponential decay may ultimately occur. The limitations are set by probe resolution and signal-to-noise constraints inherent to the smallest resolved scales. 
Further advances in nanoscale sensing will be required to establish the asymptotic form of the spectrum deep in the viscous range.

\begin{acknowledgments}
We gratefully acknowledge Shikha Shikha for her assistance with conventional hot-wire manufacturing. This work was made possible through
Office of Naval Research Grant (N00014-22-1-2038).
\end{acknowledgments}


\bibliography{references}

\end{document}